\documentclass[12pt]{iopart}

\usepackage{iopams}  %additional characters
\usepackage{citesort}  %for proper citations
\usepackage{graphicx}  %for figures
\usepackage{xcolor}    %for background color
\definecolor{brightgray}{gray}{0.7}
\definecolor{bbrightgray}{gray}{0.95}
\usepackage{comment}    %to comment large parts

\newcommand{\norm}[1]{\left|\left|#1\right|\right|}
\newcommand{\unit}{1\!\!1}
\newcommand{\Ceil}[1]{\left\lfloor #1 \right\rfloor}
\newcommand{\sgn}{\mathrm{sgn}}
\newcommand{\mod}{\mathrm{\ mod}}
\newcommand{\perm}{\mathrm{perm}}
\newcommand{\abs}[1]{\left| #1 \right|} % for absolute value

\newcommand{\gb}[1]{\colorbox{brightgray}{#1}}
\newcommand{\gbb}[1]{\colorbox{bbrightgray}{#1}}

\begin{document}
%==================================================================================
%==================================================================================
%================================TITLE=============================================
%==================================================================================
%==================================================================================

\title[Many-body quantum interference on hypercubes]{Many-body quantum interference on hypercubes}

\author{Christoph Dittel, Robert Keil and Gregor Weihs}

\address{Institut f\"ur Experimentalphysik, Universit\"at Innsbruck, Technikerstra{\ss}e 25d, A-6020 Innsbruck, Austria}
\ead{christoph.dittel@uibk.ac.at}

\begin{abstract}
Beyond the regime of distinguishable particles, many-body quantum interferences influence quantum transport in an intricate manner. However, symmetries of the single-particle transformation matrix alleviate this complexity and even allow the analytic formulation of suppression laws, which predict final states to occur with a vanishing probability due to total destructive interference. Here we investigate the symmetries of hypercube graphs and their generalizations with arbitrary identical subgraphs on all vertices. We find that initial many-particle states, which are invariant under self-inverse symmetries of the hypercube, lead to a large number of suppressed final states. The condition for suppression is determined solely by the initial symmetry, while the fraction of suppressed states is given by the number of independent symmetries of the initial state. Our findings reveal new insights into particle statistics for ensembles of indistinguishable bosons and fermions and may represent a first step towards many-particle quantum protocols in higher-dimensional structures.

\end{abstract}

\vspace{2pc}
%\submitto{\NJP}
\maketitle

%==================================================================================
%==================================================================================
%================================MAIN TEXT=========================================
%==================================================================================
%==================================================================================

%==================================================================================
%==================================================================================
%================================INTRODUCTION======================================
%==================================================================================
%==================================================================================

\section{Introduction}

Quantum transports of single particles on discrete graphs, also referred to as continuous-time quantum walks, are governed by the interference of all possible pathways as provided by the graph structure. This wavelike interference can be harnessed as an algorithmic tool in quantum computation \cite{Childs-EA-2003}, for excitation transfer in spin chains \cite{Bose-QC-2003,Yao-SA-2012,Sahling-ER-2015} and may even play a role in photosynthesis \cite{Engel-EW-2007,Lambert-QB-2013}.

If multiple indistinguishable particles propagate on the same graph, one has to take their exchange symmetry into account \cite{Hong-MS-1987,Liu-QI-1998,Loudon-FB-1998}, which leads to correlations between the particles \cite{Peruzzo-QW-2010}. Generally speaking, increasing particle numbers causes interference among a growing number of many-particle paths and gives rise to intricate evolution scenarios \cite{Tichy-II-2014}, as experimentally demonstrated in planar graphs \cite{Hong-MS-1987,Mattle-NC-1995,Peruzzo-QW-2010,Kaufman-TW-2014,Carolan-OE-2014}, as well as for two \cite{Poulios-QW-2014,Crespi-PS-2015} and three \cite{Spagnolo-TP-2013} particles in two dimensions.

Symmetries of the graph can have a strong influence on transport problems. For example, in the single-particle regime, it has been shown that symmetries permit perfect state transfer even through large spin chains \cite{Christandl-PS-2004,Kay-PS-2006}. In the realm of many-particle transport, the complexity of the dynamics is substantially simplified in the presence of symmetries in the unitary evolution matrix. To date, only few such symmetries have been investigated: The discrete Fourier transform \cite{Tichy-MP-2012,Crespi-SL-2016}, Sylvester matrices \cite{Crespi-SL-2015} and the $J_x$ lattice \cite{Dittel-EC-2015,Weimann-IQ-2016}. In all these cases, symmetries lead to analytic suppression laws, predicting whether or not a final particle configuration can occur. 

In this work, we consider the quantum transport of $N$ identical particles in hypercube graphs (HC) of arbitrary dimension. These highly symmetric graphs attracted much attention in the context of single-particle quantum walks \cite{Mackay-QW-2002,Moore-QW-2002,Kempe-DQ-2005,Alagic-DQ-2005,Krovi-HT-2006,Makmal-QW-2014,Makmal-QW-2016}. Here, we investigate how symmetries influence many-particle transport on HC graphs. We show analytically that the interference conditions are solely defined by the symmetries of the graph as well as of the initial state and do not necessarily depend on the particular graph structure. For bosons as well as for fermions, we derive suppression laws which predict final states with a vanishing probability of occurrence, due to destructive interference among the $N$-particle trajectories. We also show that the number of suppressed states depends strictly on how many relevant symmetries the initial state satisfies. Due to the underlying symmetries, the suppression law persists in generalized HCs with identical but arbitrary subgraphs on all vertices. This permits each node of the HC to carry internal degrees of freedom, making our approach compatible with implementations in various systems. 

This paper is structured as follows: In section~\ref{sec:preliminaries} we start with a brief introduction to many-body interference and discuss consequences of particle indistinguishability and statistical differences between bosons and fermions. A symmetric representation of the evolution scenario on HC graphs, followed by the symmetry suppression laws and an illustrative example, is given in section~\ref{sec:SuppLawBareHC}. The generalization to HC graphs with arbitrary but identical subgraphs on all vertices is shown in section~\ref{sec:GenSuppLawHC}. Finally, we discuss the results and provide an outlook towards possible experimental implementations in section~\ref{sec:discussion}. For completeness, all derivations of the discussed suppression laws are given in the appendix.

%==================================================================================
%==================================================================================
%================================PRELIMINARIES=====================================
%==================================================================================
%==================================================================================

\section{Many-particle interference}\label{sec:preliminaries}

A continuous-time quantum evolution on a graph with $n$ identical vertices is governed by a Hamiltonian $\hat{\mathcal{H}}$ with elements $\hat{\mathcal{H}}_{i,j}/\hbar$, specifying the transition rates between sites $i$ and $j$. Having explicitly the evolution of quantum states in mind, we will refer to the vertices of the graph as \textit{modes} in the remainder of the paper. An initial configuration of $N$ particles is fully described by the \textit{mode occupation list} $\bi{r}=(r_1,\dots,r_n)$, where each element $r_j$ specifies the number of particles in mode $j$. For convenience, particle states can also be expressed by the \textit{mode assignment list} 
\begin{eqnarray}
\bi{d}(\bi{r})=(d_{1}(\bi{r}),....,d_{N}(\bi{r})),
\end{eqnarray}
with $d_j(\bi{r})$ denoting the mode number occupied by the $j$-th particle \cite{Tichy-MP-2012,Tichy-II-2014}. A final state after projective measurement is similarly denoted by $\bi{s}=(s_1,\dots,s_n)$ and $\bi{d}(\bi{s})=(d_{1}(\bi{s}),....,d_{N}(\bi{s}))$. 

After an action of the Hamiltonian for some time $t$, a single-particle state undergoes a unitary transformation $\hat{U}=\exp(\rmi\hat{\mathcal{H}} t/\hbar)$. For the transition of many-particle states, however, all possibilities to distribute the particles among the final modes contribute. Particle indistinguishability then requires a coherent sum over all many-particle paths in the calculation of the transition probability~\cite{Mayer-CS-2011,Tichy-MP-2012,Tichy-II-2014,Crespi-SL-2015},
\begin{eqnarray}\label{eq:P}
 P_{\mathrm{B/F}}(\bi{r},\bi{s},\hat{U})=\frac{\prod_k\ s_k!}{\prod_l\ r_l!}\left|\sum_{\bsigma\in S_{\bi{d}(\bi{s})}} \sgn_{\mathrm{B/F}}(\bsigma)\prod_{j=1}^N\ \hat{U}_{d_j(\bi{r}),\sigma_j} \right|^2 ,
\end{eqnarray}
where the sum runs over the set of all possible permutations $S_{\bi{d}(\bi{s})}$ of the final mode assignment list $\bi{d}(\bi{s})$. Due to the bosonic (fermionic) (anti-)commutation-relation, one has $\sgn_{\mathrm{B}}(\bsigma)=1$ and $\sgn_{\mathrm{F}}(\bsigma)=\sgn(\bsigma)$. It must be emphasized, that the number of particles for fermions is restricted to $N\leq n$ due to Pauli's principle \cite{Pauli-EP-1964}, whereas multiple bosons are allowed to occupy the same mode. By defining the matrix $M_{j,k}\equiv U_{d_j(\bi{r}),d_k(\bi{s})}$, which contains all rows and columns of $\hat{U}$, corresponding to occupied initial and final modes, respectively, the transition probabilities
\begin{eqnarray}\label{eq:PB}
P_{\mathrm{B}}(\bi{r},\bi{s},\hat{U})=\frac{1}{\prod_k r_k!\ s_k!}\abs{\perm(M)}^2
\end{eqnarray}
and
\begin{eqnarray}\label{eq:PF}
P_{\mathrm{F}}(\bi{r},\bi{s},\hat{U})=\abs{\det(M)}^2
\end{eqnarray}
are obtained from the permanent (determinant) of transition amplitudes in a coherent manner. Thus, many-particle interference arises from the indistinguishability of the particles, that is, the absence of any which-path information in the process. For comparison, the transition probability for distinguishable particles~\cite{Tichy-MP-2012,Tichy-II-2014}
\begin{eqnarray}\label{eq:PD}
P_{\mathrm{Dist}}(\bi{r},\bi{s},\hat{U})=\frac{1}{\prod_k r_k!\ s_k!}\perm(\abs{M})^2,
\end{eqnarray}
contains no phase dependence and is, thus, not affected by many-body interference.

Essentially, symmetries simplify the complex interplay in many-body transitions and result in a structured ordering, which individually affects the arising interferences for bosons and fermions. A closer look at the calculation of transition probabilities according to (\ref{eq:PB}) and (\ref{eq:PF}) reveals, that symmetries of the unitary matrix $\hat{U}$ will transfer into $M$ for appropriate choices of the initial and the final state. Consequently, the resulting symmetries in $M$ strongly affect the outcome in (\ref{eq:PB}) and (\ref{eq:PF})~\cite{Tichy-MP-2012}. In particular, total destructive interference will occur for certain combinations of initial and final states, which is discussed for HC graphs in the following.

%==================================================================================
%==================================================================================
%================================BARE SUPPRESSION LAW==============================
%==================================================================================
%==================================================================================

\section{Symmetry suppression law on HC graphs}\label{sec:SuppLawBareHC}

%=============================HC UNITARY===========================================
\subsection{HC unitary}

\begin{figure}[t]
\centering
\includegraphics[width=10cm]{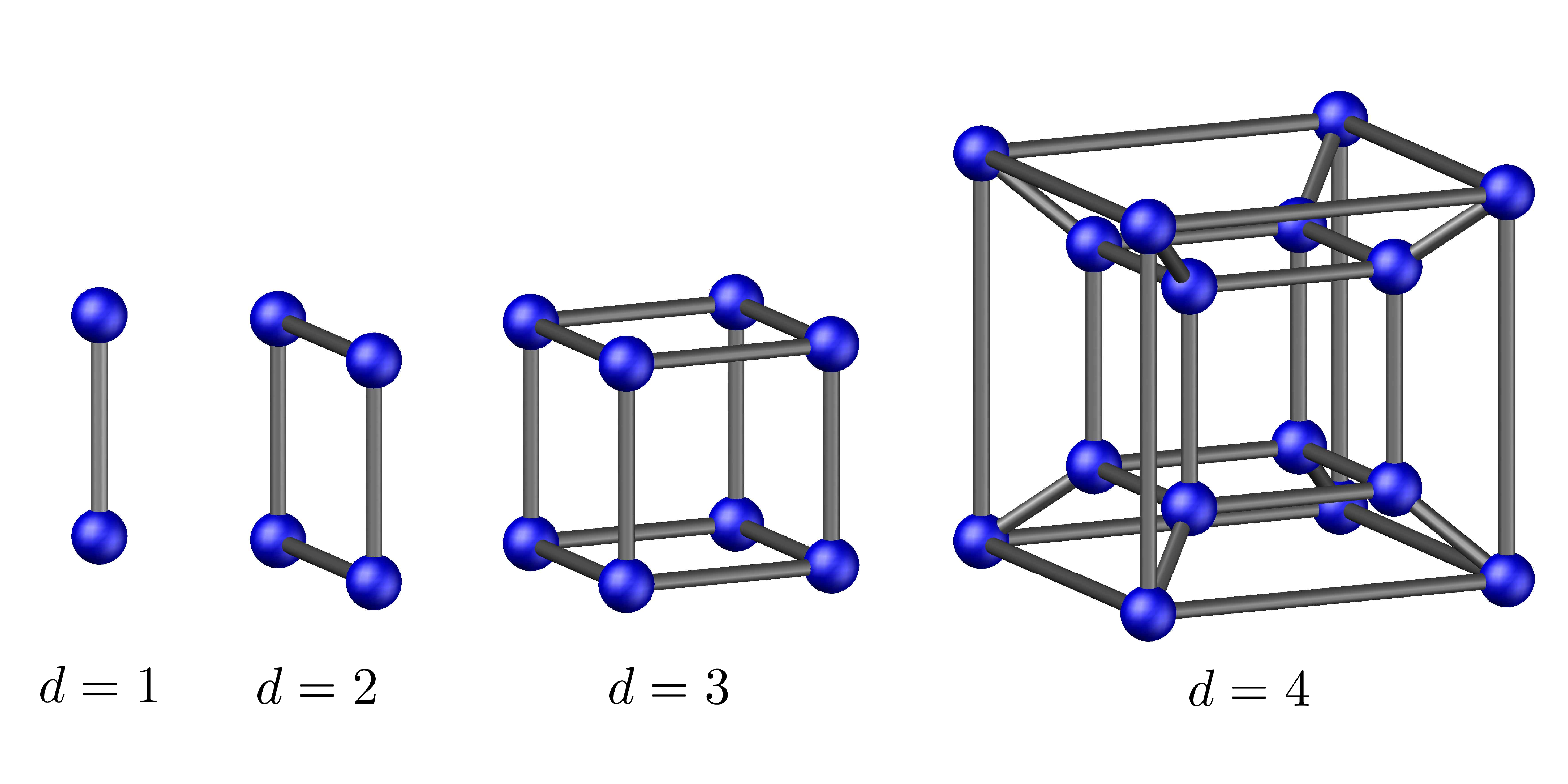}
\caption{Schematic illustration of HC graphs up to four dimensions. All $n=2^d$ modes (blue balls) are equally coupled (silver connections) to $d$ neighbouring modes.}
\label{fig:HC3d}
\end{figure}

The $d$-dimensional hypercube consists of $n=2^d$ modes, each connected to $d$ neighbours as illustrated in Figure~\ref{fig:HC3d}.  
Note that this system may be seen as a generalisation of a $2$-site graph in $d$ dimensions. Let $\hat{\mathcal{H}}_{i,j}/\hbar=\kappa$ be the transition rate for neighbouring sites $i$ and $j$. For an evolution time $t=\pi/(4\kappa)$, the unitary transformation matrix reads
\begin{eqnarray} \label{eq:U}
\hat{U}=\frac{1}{\sqrt{n}}\left(\begin{array}{cc} 1&\rmi\\ \rmi&1\end{array}\right)^{\otimes\ d}.
\end{eqnarray}
It contains all single-particle transformation amplitudes and is the basis for calculating many-particle amplitudes according to~(\ref{eq:PB})-(\ref{eq:PD}). Due to the tensor power in (\ref{eq:U}), the unitary clearly exhibits stepwise symmetries. Pertaining to equation~(\ref{eq:P}), an explicit expression for the element $\hat{U}_{j,k}$ is required to make the importance of symmetries apparent. To obtain such an expression, sign-bookkeeping instruments are introduced: We define segmentation parameters $p\in\{2,4,8,\dots,n\}$ and their corresponding Rademacher functions \cite{Rademacher-ES-1922}
\begin{eqnarray}\label{eq:x}
x(j,p)=(-1)^{ \Ceil{\frac{p(j-1)}{n}}}
\end{eqnarray}
where the Gaussian brackets $\Ceil{\alpha/\beta}$ evaluate the quotient of the Euclidean division $\alpha/\beta$. The Rademacher functions, exemplified for $d=3$ in the first three rows of table~\ref{tab:x}, assign the value $1$ ($-1$) to modes $j\in\{1,\dots,n\}$ as determined
by the segmentation value $p$ in a symmetric and stepwise fashion, such that all modes are clearly grouped into two subsets of the same size. 
\begin{table}
\caption{\label{tab:x}Rademacher functions and Walsh functions for the three dimensional HC. To guide the eye, the numbers $1$ and $-1$ are shaded in dark gray and light gray, respectively.}

\begin{indented}
\lineup
\item[]\begin{tabular}{@{}*{10}{l}}
\br
 & &\centre{8}{Mode number $j$}\\
Rademacher and Walsh Functions& & 1& 2& 3& 4& 5& 6& 7& 8\\
\mr
$x(j,2)$&&\gb{1}&\gb{1}&\gb{1}&\gb{1}&\gbb{\-1}&\gbb{\-1}&\gbb{\-1}&\gbb{\-1}\cr
$x(j,4)$&&\gb{1}&\gb{1}&\gbb{\-1}&\gbb{\-1}&\gb{1}&\gb{1}&\gbb{\-1}&\gbb{\-1}\cr
$x(j,8)$&&\gb{1}&\gbb{\-1}&\gb{1}&\gbb{\-1}&\gb{1}&\gbb{\-1}&\gb{1}&\gbb{\-1}\cr
\mr
$\mathcal{A}(j,(2,4))=x(j,2)\ x(j,4)$&&\gb{1}&\gb{1}&\gbb{\-1}&\gbb{\-1}&\gbb{\-1}& \gbb{\-1}&\gb{1}&\gb{1}\cr
$\mathcal{A}(j,(2,8))=x(j,2)\ x(j,8)$&&\gb{1}&\gbb{\-1}&\gb{1}&\gbb{\-1}&\gbb{\-1}& \gb{1}&\gbb{\-1}&\gb{1}\cr
$\mathcal{A}(j,(4,8))=x(j,4)\ x(j,8)$&&\gb{1}&\gbb{\-1}&\gbb{\-1}&\gb{1}&\gb{1}&\gbb{\-1}&\gbb{\-1}&\gb{1}\cr
\mr
$\mathcal{A}(j,(2,4,8))=x(j,2)\ x(j,4)\ x(j,8)$&&\gb{1}&\gbb{\-1}&\gbb{\-1}&\gb{1}&\gbb{\-1}&\gb{1}&\gb{1}&\gbb{\-1}\cr
\br
\end{tabular}
\end{indented}
\end{table}

Returning to unitary~(\ref{eq:U}), each element differs by a phase shift, equalling an integer multiple of $\pi/2$. The actual phase of element $\hat{U}_{i,j}$ can be calculated by means of the segmentations $p$, accounting for the accumulation of phases in the $i$-th row and $j$-th column. Thus, the elements of the unitary can be written as
\begin{eqnarray}\label{eq:Uelement}
\hat{U}_{j,k}=\frac{1}{\sqrt{n}}\ \exp\left(\rmi\ \frac{\pi}{4}\left[d-\sum_{l=1}^{d}\ x(j,2^l)\ x(k,2^l)\right]\right)
\end{eqnarray}
as shown in~\ref{aqq:Uelements}. The simple phase relation in this expression is of advantage when evaluating interference contributions for the calculation of transition probabilities in~(\ref{eq:P}).

Note that the unitary~(\ref{eq:U}) is equivalent to the one considered in~\cite{Crespi-SL-2015} except for local phase operations. Consequently, suppression laws have also been discovered in Sylvester interferometers for specific initial states with particle numbers in powers of two. As we show in the following, however, by making use of the HC symmetries, one can prove suppression laws for a much larger class of particle configurations.

%=============================SYMMETRY OPERATIONS==================================
\subsection{Symmetry operations}
\begin{figure}[t]
\centering
\includegraphics[width=12cm]{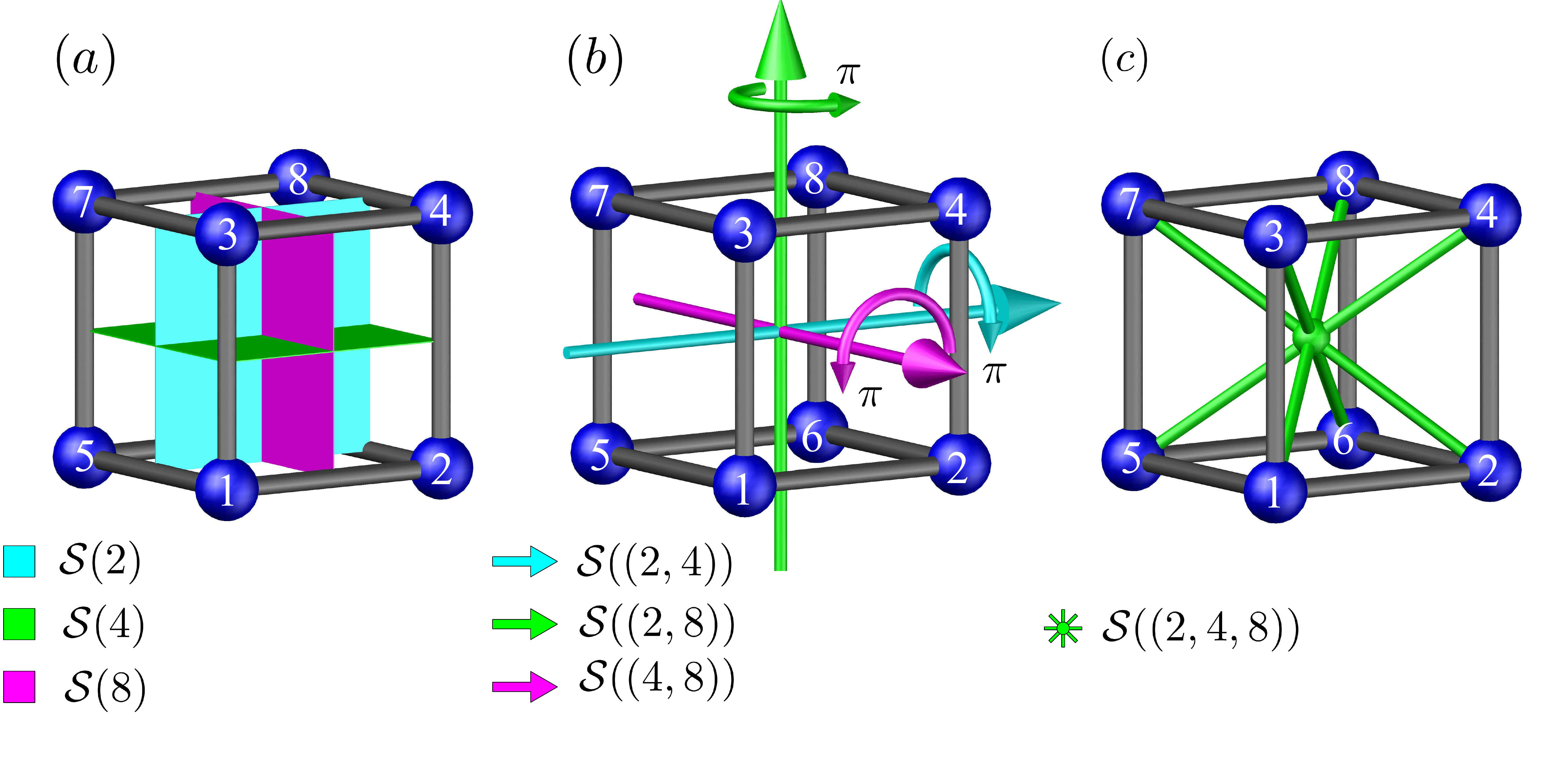}
\caption{Illustration of self-inverse symmetry operations on the three dimensional HC. The operators $\mathcal{S}(2)$, $\mathcal{S}(4)$ and $\mathcal{S}(8)$ act as plane mirror symmetry operations, indicated on cube $(a)$. Operators with $\norm{\bi{p}}=2$, $\mathcal{S}((2,4))$, $\mathcal{S}((2,8))$ and $\mathcal{S}((4,8))$, are illustrated on cube $(b)$ and act as $\pi$-rotations. Cube $(c)$ shows the action of  $\mathcal{S}((2,4,8))$, which corresponds to the point reflection with respect to the center.}
\label{fig:SymOpD3}
\end{figure}

We consider the self-inverse symmetry operations 
\begin{eqnarray}\label{eq:S}
\mathcal{S}(p)=\unit^{\otimes \log_2(p/2)}\otimes \sigma_x \otimes \unit^{\otimes \log_2(n/p)},
\end{eqnarray}
where $\unit$ denotes the $2\times 2$ identity matrix and $\sigma_x$ the $2\times 2$ Pauli spin-x-operator. These symmetry operators are mutually commuting and act on mode occupation lists as $p/2$ transpositions of all $p$ segments, according to
\begin{eqnarray}
\left[\mathcal{S}(p)\ \bi{r}\right]_j= \sum_{k=1}^n \mathcal{S}(p)_{j,k}\ r_k=r_{j+x(j,p)\frac{n}{p}},
\end{eqnarray}
which can be interpreted as a reflection in dimension $\log_2(p)$.
Moreover, a consecutive action of symmetry operations with different $p$ values, where each operator acts at most once, is denoted by 
\begin{eqnarray}\label{eq:SymOpList}
\prod_{k=1}^{\norm{\bi{p}}} \mathcal{S}(p_k)\equiv \mathcal{S}(\bi{p}),
\end{eqnarray}
with $\bi{p}=(p_1,p_2,\dots)$ and $\norm{\bi{p}}$ denoting the number of elements in $\bi{p}$. Thus, the total number of possible symmetry operations for the $d$ dimensional HC corresponds to the number of subsets of $(2,4,\dots,2^n)$, omitting the empty list. These $2^d-1$ self-inverse symmetries are key to the classification of the arising many-particle interferences. Figure~\ref{fig:SymOpD3} demonstrates the case of a three dimensional HC, where all self-inverse operations are assigned to familiar symmetries. For compositions of symmetry-operators, it is convenient to introduce Walsh functions \cite{Walsh-CS-1923}
\begin{eqnarray}\label{eq:partition}
\mathcal{A}(j,\bi{p})\equiv\prod_{m=1}^{\norm{\bi{p}}}\ x(j,p_m),
\end{eqnarray}
which assign $1$~$(-1)$ to all modes as governed by the symmetry set $\bi{p}$, thereby partitioning all modes into two complementary subsets $\mathcal{P}(\bi{p})=\{j\in \{1,\dots,n\}\ |\ \mathcal{A}(j,\bi{p})=-1\}$ and $\bar{\mathcal{P}}(\bi{p})=\{j\in \{1,\dots,n\}\ |\ \mathcal{A}(j,\bi{p})=1\}$. For the sake of completeness, the lower rows in table~\ref{tab:x} list the partitionings for any composite set $\bi{p}$ on the three dimensional HC.

%=============================SUPPRESSION LAW======================================
\subsection{Suppression law}
For initial states, being invariant under the symmetry operation
\begin{eqnarray}\label{eq:invariant}
\mathcal{S}(\bi{p})\ \bi{r}=\bi{r},
\end{eqnarray} 
the symmetry of the unitary is in some way preserved in matrix $M$. Note that an initial state can be invariant under various symmetries which may be related to each other. In this regard, we denote the number of independent symmetries of an initial state by $\eta$, being the  minimal number of symmetry operations with which one can compose all symmetry operations the initial state is invariant under\footnote{Recalling that the symmetry operators $\mathcal{S}(\bi{p})$ are self-inverse and mutually commute, we define the set $\Gamma=\{\bi{p}\ |\ \mathcal{S}(\bi{p})\ \bi{r}=\bi{r}\}$ and all possible sets $\Lambda_k \subseteq \Gamma\ :\ \forall\bi{p}\in\Gamma\ \exists\ T\subseteq \Lambda_k\ :\ \prod_{\bi{p}_j\in T}\mathcal{S}(\bi{p}_j)=\mathcal{S}(\bi{p}) $, which generate all symmetry operations the initial state is invariant under. Then, the number of independent symmetries of $\bi{r}$ is given by $\eta=\min\{\abs{\Lambda_1},\abs{\Lambda_2},\dots\}$. For example, if $\bi{r}$ is invariant under $\mathcal{S}(2)$, $\mathcal{S}(8)$ and $\mathcal{S}((2,8))$, then $\Lambda_1=\{2,8,(2,8)\}$, $\Lambda_2=\{2,8\}$, $\Lambda_3=\{2,(2,8)\}$, $\Lambda_4=\{8,(2,8)\}$ and accordingly $\eta=2$.} . In order to satisfy condition~(\ref{eq:invariant}), particles must be arranged in a configuration which is invariant under all independent symmetries. Therefore, the particle number is restricted to $N=z\cdot 2^{\eta}$ with $z \in \mathbb{N}$.

Under these conditions, the underlying symmetry can lead to a vanishing transition probability for certain final states. This arises from the $N$-particle transition amplitudes of~(\ref{eq:P}) summing up in pairs of permutations, which cancel out due to their opposite phases. The partitioning of all $n$ modes according to the Walsh functions in~(\ref{eq:partition}) then allows the formulation of suppression laws, that predict which final states are suppressed:

\textbf{Bosons:} \textit{For an initial state $\bi{r}$ of $N$ indistinguishable bosons, which is invariant under the symmetry operation $\mathcal{S}(\bi{p})$, such that $N$ must be even, all final states $\bi{s}$ with an odd number of particles in the subset of modes $k$ for which $\mathcal{A}(k,\bi{p})=-1$, are suppressed, i.e.,}
\numparts
\begin{eqnarray}\label{eq:supplaw1}
\prod_{j=1}^N \mathcal{A}(d_j(\bi{s}),\bi{p})=-1\ \ \Rightarrow\ \ P_\mathrm{B}(\bi{r},\bi{s},\hat{U})=0.
\end{eqnarray}

\textbf{Fermions:} \textit{For an initial state $\bi{r}$ of $N$ indistinguishable fermions, which is invariant under the symmetry operation $\mathcal{S}(\bi{p})$, such that $N$ must be even, all final states $\bi{s}$, that do not have exactly $N/2$ particles in the subset of modes $k$ for which $\mathcal{A}(k,\bi{p})=-1$, are suppressed, i.e.,}
\begin{eqnarray}\label{eq:supplaw2}
\sum_{j=1}^N \mathcal{A}(d_j(\bi{s}),\bi{p})\neq 0\ \ \Rightarrow\ \ P_\mathrm{F}(\bi{r},\bi{s},\hat{U})=0.
\end{eqnarray}
\endnumparts

The proofs of these suppression laws are given in \ref{aqq:SupLawProof}. The direct connection of the involved symmetries to the condition for total destructive interference is immediately apparent by comparing the symmetry operation~(\ref{eq:SymOpList}) with the partitioning~(\ref{eq:partition}). These symmetries simplify the underlying complexity of many-particle evolutions in a demonstrative way (see section~\ref{sec:bosonexample} for an explicit example). Even if no statement about probabilities of non-vanishing events is made by the suppression laws (\ref{eq:supplaw1}) and~(\ref{eq:supplaw2}) themselves, they are obtained purely analytically, thereby permitting a prediction which final states must be suppressed with little computational overhead.

A deeper comparison of the suppression laws for bosons and fermions reveals an interesting connection: First of all, fermions feature a much more general condition for suppression as compared to bosons, such that the set of allowed configurations is more restricted for the former. Whether the suppressed boson states are also forbidden for fermions depends on the parity of $N/2$: For particle numbers satisfying $\mod[N,4]=0$ the two suppression conditions coincide, that is, states with an odd number of particles occupying the relevant set of modes are suppressed for both types of particles. However, for $\mod[N,4]=2$, the conditions are complementary to each other. A similar effect was discovered in the studies of the discrete Fourier transform in~\cite{Tichy-MP-2012}, where the particle number also determines whether or not the suppression law for many-fermion states resembles the one for bosons.

%=============================SUPPRESSION RATIO====================================
\subsection{Suppression ratio}

In order to highlight the extent of many-body destructive interference on the considered graphs, the suppression ratios $\mathcal{N}_{\mathrm{supp}}/\mathcal{N}_{\mathrm{all}}$ for the derived suppression laws~(\ref{eq:supplaw1}) and~(\ref{eq:supplaw2}) are discussed in the following. Here $\mathcal{N}_{\mathrm{all}}$ denotes the total number of final states and $\mathcal{N}_{\mathrm{supp}}$ the predicted number of suppressed states. 

According to the suppression law~(\ref{eq:supplaw1}), for bosonic initial states, which are invariant under a single independent symmetry operation ($\eta=1$), one expects the suppression of about half of all possible final states. To respect common suppressed states of different independent symmetries, the intersection of their respective subsets  $\mathcal{P}(\bi{p})$ and $\mathcal{P}(\bi{p}')$ has to be considered. From the definition of the Walsh functions in~(\ref{eq:partition}), one finds $\sum_{j=1}^n\mathcal{A}(j,\bi{p})\mathcal{A}(j,\bi{p}')=0$ for $\bi{p}\neq\bi{p}'$ (see for example table~\ref{tab:x}). Recalling that $\mathcal{P}(\bi{p})$ and $\bar{\mathcal{P}}(\bi{p})$ are complementary sets of equal size, this implies that subsets corresponding to different symmetry operations share half of their elements, i.e. $|\mathcal{P}(\bi{p})\cap \mathcal{P}(\bi{p}')|=|\mathcal{P}(\bi{p})|/2=n/4$. Thus, two different independent symmetries have about half of all suppressed final states in common. It follows, that the suppression ratio for those bosonic initial states can be approximated for large $n$ by
\begin{eqnarray}\label{eq:suppratioBosons}
\frac{\mathcal{N}^{\mathrm{B}}_{\mathrm{supp}}}{\mathcal{N}^{\mathrm{B}}_{\mathrm{all}}}\approx\sum_{j=1}^\eta \frac{1}{2^j}=1-\frac{1}{2^\eta}.
\end{eqnarray}

\begin{figure}[t]
\centering
\includegraphics[width=9.5cm]{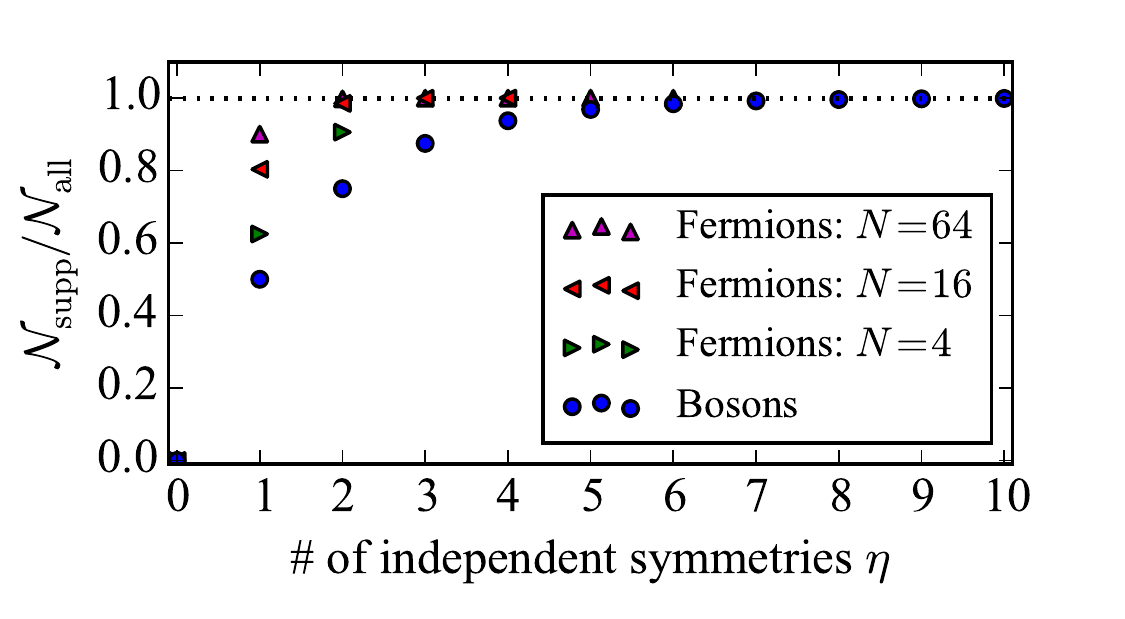}
\caption{Estimated suppression ratio as a function of the number of independent symmetries an initial state is invariant under. Blue circles show the suppression ratios for bosons according to equation~(\ref{eq:suppratioBosons}) whereas magenta, red and green triangles show the suppression ratios for $N=64$, $N=16$ and $N=4$ fermions, respectively, according to equation~(\ref{eq:suppratioFermions}). Note that in the case of fermions, the maximal number of independent symmetries is bounded by the particle number.}
\label{fig:SuppRatio}
\end{figure} 
 
In the case of fermions, for each independent symmetry, the suppression law~(\ref{eq:supplaw2}) forces all particles to distribute equally among the two complementary subsets, regardless to the occupation within each individual subset. Furthermore, subsets corresponding to different symmetries share one half of their elements. Thus, for $\eta$ independent symmetries the total number of subsets, among which particles are forced to distribute equally, is given by $2^\eta$. Accordingly, the number of states, which are not covered by the suppression law~(\ref{eq:supplaw2}), is given by ${n/2^\eta\choose N/2^\eta }^{2^\eta}$. Because Pauli's principle excludes modes with multiple occupation, the total number of final states reads $\mathcal{N}_{\mathrm{all}}={n\choose N}$, resulting in
\begin{eqnarray}\label{eq:suppratioFermions}
\frac{\mathcal{N}^{\mathrm{F}}_{\mathrm{supp}}}{\mathcal{N}^{\mathrm{F}}_{\mathrm{all}}}=1-{n/2^\eta\choose N/2^\eta }^{2^\eta}{n\choose N }^{-1}
\approx 1-\frac{N!}{2^{\eta N}\left[\left(\frac{N}{2^\eta}\right)!\right]^{2^\eta}},
\end{eqnarray}
with the approximation valid for $n\gg N$. Due to symmetry reasons, fermionic initial states with particle numbers satisfying $\mod[N,4]=2$ can only be invariant under at most one independent symmetry operation. However, states with particle numbers $N=2^\eta$ can satisfy up to $\eta$ independent symmetry operations leading to a suppression ratio $\mathcal{N}^{\mathrm{F}}_{\mathrm{supp}}/\mathcal{N}^{\mathrm{F}}_{\mathrm{all}}\approx 1-N!/N^N$. For comparison, Figure~\ref{fig:SuppRatio} illustrates the suppression ratios for bosons and fermions in the limit $n\gg N$. It is evident how, for a fixed number of independent symmetries, more fermionic states are suppressed than for bosons and that fermionic suppression gets increasingly restrictive for growing particle numbers.

%==================================================================================
%================================EXAMPLE===========================================
%==================================================================================

\subsection{Suppression of bosonic states on the cube}\label{sec:bosonexample}

\begin{figure}[t]
\centering
\includegraphics[width=\textwidth]{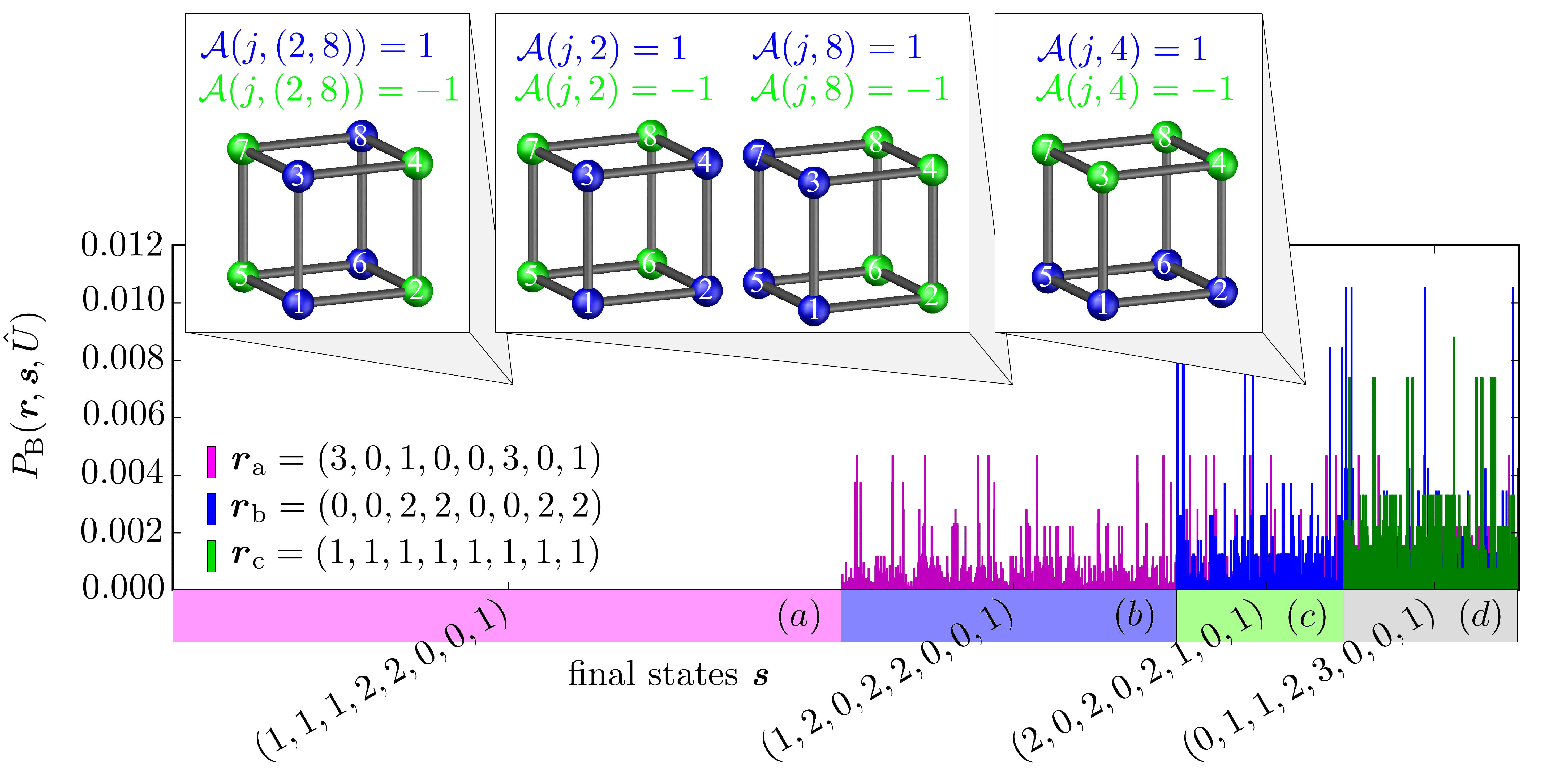}
\caption{Transition probabilities for three symmetric initial states with $N=8$ bosons on the three dimensional HC. On the horizontal axis, all possible final states are grouped into four sets. Set $(a)$, coloured light magenta, contains states which must be suppressed for $\bi{p}=(2,8)$. In set $(b)$, coloured light blue, suppressed states for $p=2$ and $p=8$ are grouped, excluding those already contained in $(a)$. States which must be suppressed for $p=4$ but not contained in $(a)$ or $(b)$, are grouped in set $(c)$, coloured light green. All remaining states are contained in set $(d)$, coloured gray. The ordering of states within each set is chosen arbitrarily. Magenta, blue and green bars indicate the transition probabilities for initial states $\bi{r}_\mathrm{a}$, $\bi{r}_\mathrm{b}$ and $\bi{r}_\mathrm{c}$, respectively. The insets visualize the corresponding partitionings. For clarity, only four of 6435 final states are labeled, representing the characteristics of states in sets $(a)$-$(d)$, respectively.}
\label{fig:PBexample}
\end{figure}

To exemplify the above suppression law, a scattering scenario of $N=8$ bosons on the HC for $d=3$ dimensions is considered. In total there are ${N+n-1\choose N}=6435$ possibilities to distribute the particles over the $n=8$ modes, whereof three initial states are chosen for discussion. For these states, Figure~\ref{fig:PBexample} shows the transition probabilities for all possible final states $\bi{s}$. 
The initial state $\bi{r}_\mathrm{a}=(3,0,1,0,0,3,0,1)$ is only invariant under the operation $\mathcal{S}((2,8))$, causing the suppression of all final states with an odd number of bosons in the subset of modes $k$ for which $\mathcal{A}(k,(2,8))=-1$. These states are grouped in set~$(a)$ of Figure~\ref{fig:PBexample}, where the corresponding partitioning is visualized in the first inset. One representative of those is the final state $\bi{s}=(1,1,1,2,2,0,0,1)$ containing three (five) bosons in modes $k$ for which $\mathcal{A}(k,(2,8))=1$ ($-1$).

The second initial state under consideration, $\bi{r}_\mathrm{b}=(0,0,2,2,0,0,2,2)$, is invariant under the operations $\mathcal{S}(2)$ and $\mathcal{S}(8)$ and accordingly also under $\mathcal{S}((2,8))$. Thus, in addition to the suppressed final states for $\bi{r}_\mathrm{a}$, states with an odd number of bosons in the subset of modes $k$ for which $\mathcal{A}(k,2)=-1$ and an odd number in the subset of modes with $\mathcal{A}(k,8)=-1$ must also be suppressed. This is shown in set~$(b)$ in Figure~\ref{fig:PBexample}. One representative of this set is the final state $\bi{s}=(1,2,0,2,2,0,0,1)$, exhibiting five (three) bosons in modes $k$ for which $\mathcal{A}(k,2)=1$ ($-1$) and three (five) bosons in modes $k$ for which $\mathcal{A}(k,8)=1$ ($-1$). For clarity, the second inset illustrates the partitionings for $p=2$ and $p=8$. 

In this scattering scenario, the initial states with the largest number of invariant symmetry operations is given by $\bi{r}_\mathrm{c}=(1,1,1,1,1,1,1,1)$. Therefore,
states with an odd number of bosons in the subset of modes $k$ for which $\mathcal{A}(k,4)=-1$ must also be suppressed in addition to all states already suppressed for $\bi{r}_\mathrm{b}$. In Figure~\ref{fig:PBexample}, these states are grouped in set~$(c)$, excluding those, already contained in set $(a)$ and $(b)$. The corresponding inset shows the partitioning for $p=4$. As a representative, the state $\bi{s}=(2,0,2,0,2,1,0,1)$ is chosen, containing five (three) bosons in modes $k$ for which $\mathcal{A}(k,4)=1$ ($-1$).
All remaining final states that are not covered by the suppression law for any $\bi{p}$, like $\bi{s}=(0,1,1,2,3,0,0,1)$, are grouped in set~$(d)$.

Here it becomes apparent that the symmetry of many-body initial states determines a well structured appearance of total destructive interference on HC graphs. It also highlights the interdependence of different symmetry operations. For example, the initial state $\bi{r}_\mathrm{b}$ is invariant under two independent symmetry operations, namely  $\mathcal{S}(2)$ and $\mathcal{S}(8)$. From this follows that the state must be invariant under a composition of both operators as well. Thus final states, which are suppressed for $\bi{r}_\mathrm{a}$ (invariant under $\mathcal{S}((2,8))$) and covered by the suppression law, must also be suppressed for $\bi{r}_\mathrm{b}$. However, since $\bi{r}_\mathrm{a}$ is only invariant under $\mathcal{S}((2,8))$, states which are suppressed for $\bi{r}_\mathrm{b}$ are not necessarily suppressed for $\bi{r}_\mathrm{a}$. 

Figure~\ref{fig:PBexample} also demonstrates the approximated suppression ratio~(\ref{eq:suppratioBosons}) in dependence on the number of independent symmetries an initial state is invariant under. Since $\bi{r}_\mathrm{a}$ is only invariant under $\mathcal{S}((2,8))$, half of all final states are expected to be suppressed, as apparent by the size of set~$(a)$ in Figure~\ref{fig:PBexample}. For each further independent symmetry, one clearly sees the additional suppression of half of the remaining final states from the size reduction of the other sets.

%==================================================================================
%==================================================================================
%================================GENERAL SUPPRESSION LAW===========================
%==================================================================================
%==================================================================================
\section{Generalization to arbitrary subunitaries}\label{sec:GenSuppLawHC}

The derivations of the discussed HC suppression laws (see appendix~\ref{aqq:SupLawProof} for details) are exclusively based on symmetric connections between unitary elements with respect to predefined symmetry operations. Thus, the suppression laws are not restricted to HC graphs as illustrated in Figure~\ref{fig:HC3d} and can be made applicable to different types of unitaries. This requires a proper redefinition of the symmetry operations~(\ref{eq:S}) such that the symmetric connections between related unitary elements remain. 

\subsection{Modifications of the symmetry conditions}

Now, in order to generalize the derived suppression laws, $d$-dimensional HC graphs are considered where each vertex consists of the same but arbitrary subgraph. Each subgraph then contains $m$ modes. Besides the coupling within each subgraph, each mode must be equally coupled to its $d$ identical counterparts in the predetermined HC ordering. This generalizes the model considerably since the HC-vertices are allowed to have diverse internal degrees of freedom. Such scenarios can be described by means of a $m\times m$ subunitary $\hat{A}$, where the overall unitary
\begin{eqnarray}\label{eq:unitarygen}
\hat{\mathcal{U}}=\frac{1}{\sqrt{2^d}}\ \hat{A}\otimes \left(\begin{array}{cc} 1&\rmi \\ \rmi &1\end{array}\right)^{\otimes\ d}
\end{eqnarray}
consists of $2^{2d}$ subunitaries. In order to retain the HC symmetries for segmentations $p\in\{2,4,8,\dots,2^d\}$, the symmetry operations~(\ref{eq:S}) are generalised to
\begin{eqnarray}\label{eq:SymGen}
\mathcal{S}(p)=\unit^{\otimes \log_2(p/2)}\otimes \Sigma_x \otimes \unit^{\otimes \log_2(2^d/p)}
\end{eqnarray}
by replacing the Pauli spin-x-operator with the exchange-operator $\Sigma_x$ of dimension $2m\times 2m$, reading
\begin{eqnarray}
\Sigma_x=\left(\begin{array}{cc}
\hat{0}_{m\times m}& \unit_{m\times m} \\ 
\unit_{m\times m}& \hat{0}_{m\times m}
\end{array}\right).
\end{eqnarray}
Here $\hat{0}_{m\times m}$ denotes the $m\times m$ null matrix and $\unit_{m\times m}$ the $m\times m$ identity matrix. These modifications are sufficient in order to make the symmetry suppression laws~(\ref{eq:supplaw1}) and~(\ref{eq:supplaw2}) applicable to any unitary given by~(\ref{eq:unitarygen}). The detailed proof is given in~\ref{app:GeneralProof}.

Remarkably, for symmetry reasons, the suppression laws for these generalized graphs require the symmetric occupation of all modes for initial states. However, for final states, only the total number of particles within the subgraphs is relevant. Thus, the prediction whether or not a final state must be suppressed, is independent on its occupation within the subgraph, as will be further discussed in the following.

%==================================================================================
%================================EXAMPLE===========================================
%==================================================================================

\subsection{Bosonic interference on a generalized HC structure}
\begin{figure}[t]
\centering
\includegraphics[width=10cm]{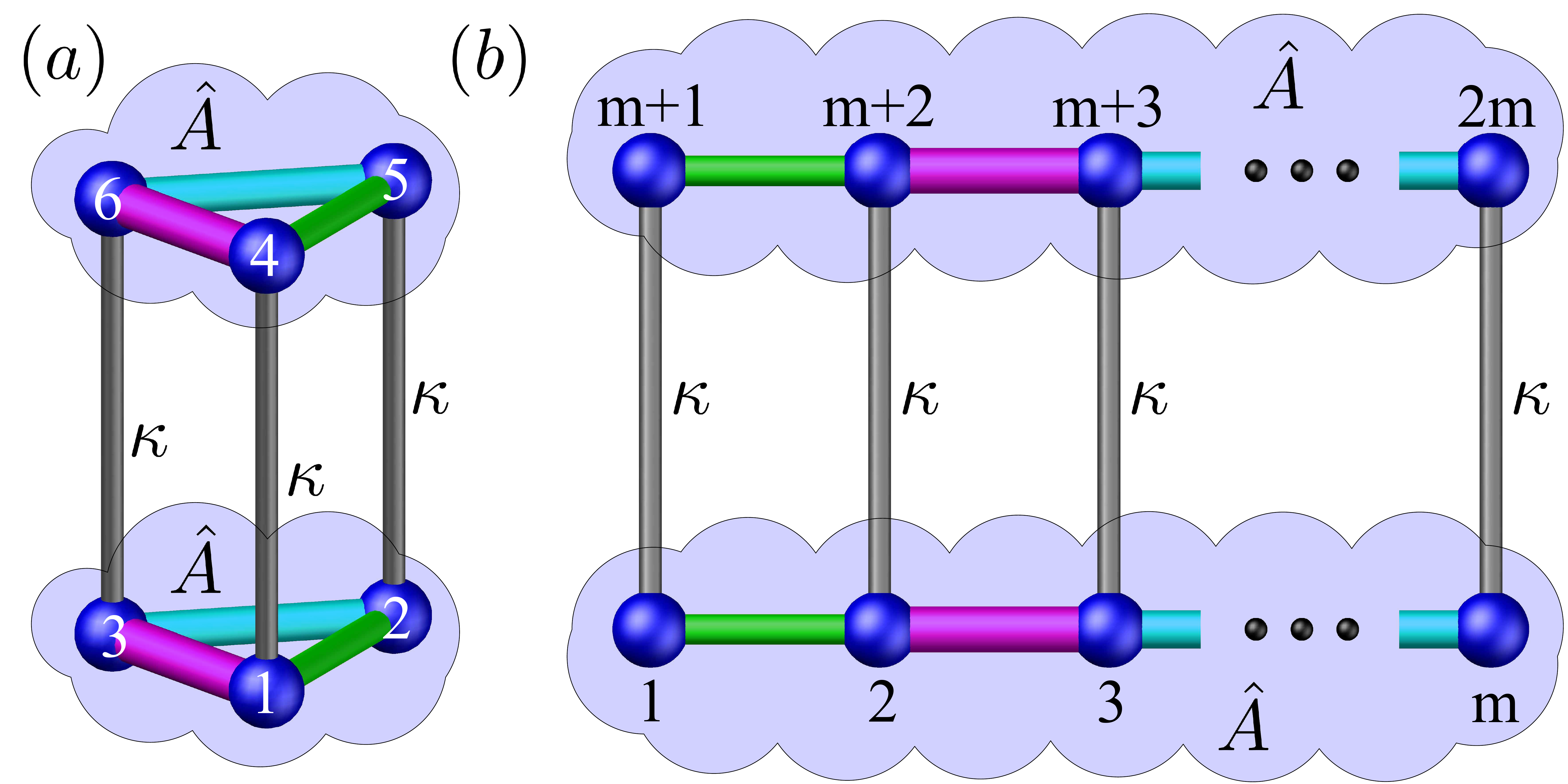}
\caption{Mode structuring for generalized one-dimensional HC graphs. Each vertex of the HC graph, framed by clouds, is composed of the same but arbitrary subgraph, specified by the subunitary $\hat{A}$. While the transition rates within the subgraph can be arbitrary (indicated by connections of different size and color) the transition rates between the subgraphs itself must equal $\kappa$. In $(a)$ the subgraph is composed of a triangular graph, whereas in $(b)$ it is composed of an arbitrary long chain with $m$ modes.}
\label{fig:Trion}
\end{figure}

In order to illustrate the generalization, bosonic state transformations in one-dimensional HC graphs are considered. Figure~\ref{fig:Trion} illustrates the mode coupling for identical subgraphs. Here, related modes of different subgraphs are equally connected by the transition rate $\kappa$, whereas transition rates within subgraphs are arbitrary but identical. 
As apparent by the unitary~(\ref{eq:unitarygen}) and by comparing Figure~\ref{fig:Trion}$(a)$ and \ref{fig:Trion}$(b)$, for the suppression law, generalized HC graphs of the same dimension are closely related, regardless of structure and size of the subgraphs. Thus, we restrict the discussion of the examples shown in Figure~\ref{fig:Trion} to the generalized graph with triangular subgraph.

Considering the mode labelling as shown in Figure~\ref{fig:Trion}$(a)$, the four-particle initial state $\bi{r}_\mathrm{a}=(2,0,0,2,0,0)$ is invariant under symmetry operation~(\ref{eq:SymGen}) for $p=2$. Then, for an evolution time $t=\pi/(4\kappa)$, the generalized suppression law predicts, that all final states with an odd number of particles on any subgraph must be suppressed. Formally, this can be seen by considering definition~(\ref{eq:partition}), which gives $\mathcal{A}(j,2)=1$ for $j\in\{1,\dots,m\}$ and $\mathcal{A}(j,2)=-1$ for $j\in\{m+1,\dots,2m\}$ with $m=3$ in the discussed example. Since each mode of a particular subgraph belongs to the same subset (i.e. their partitioning value $\mathcal{A}(j,p)$ is equal), the condition for suppression appears independently of the particle occupation within the subgraphs. Thus, for example the final states $\bi{s}=(3,0,0,0,1,0)$ and $\bi{s}=(1,1,1,0,1,0)$ must be suppressed for $\bi{r}_\mathrm{a}$. However, in agreement with the formal suppression law, they appear with a non-vanishing probability for the initial state $\bi{r}_\mathrm{b}=(2,0,0,1,1,0)$. This highlights that the condition for suppression strongly depends on the initial particle occupation within the subgraphs as determined by the invariance under the symmetry operations~(\ref{eq:SymGen}), which $\bi{r}_\mathrm{b}$ does not fulfil.

%==================================================================================
%================================DISCUSSION========================================
%==================================================================================

\section{Discussion}\label{sec:discussion}
In this paper, we have shown that symmetries in many-particle quantum transport on HC graphs allow the formulation of analytic suppression laws, which predict final states occurring with a vanishing probability due to many-particle interference. 
Specifically, each symmetry of the initial state groups all modes into two subsets of equal size and the occupation of these subsets determines the suppression. For bosons, the suppression depends on the parity of this occupation. For fermions, on the other hand, all events with imbalanced occupation of the subsets are suppressed, giving rise to a much tighter restriction on allowed events as compared to bosons. This behaviour can be understood by regarding the HC as an extension of the 2-site coupler to higher dimensions: The forced splitting of fermions between the subsets, may be interpreted as a strong anti-bunching effect, while the distribution of bosons in sets of even parity is reminiscent of the multi-photon interference at the coupler \cite{Campos-QM-1989,Ou-OF-1999}. Moreover, the suppression law holds for HCs with arbitrary identical subgraphs on all vertices. This illustrates that it is ultimately the symmetry of the unitary which governs the suppression, not necessarily a particular graph structure. Strikingly, a similar observation has been made in single-particle quantum transport through disordered networks, where also symmetries and certain structural features, rather than the specific geometry of the graph, determine the transport efficiency \cite{Mostarda-SD-2013,Walschaers-OD-2013}.

As we have shown here for HC graphs, the emergence of symmetries in many-body state transformations of indistinguishable particles causes conceptional simplifications and allows the definite exclusion of certain outcomes. In the absence of any symmetry, however, no simple and exact examination for large many-body systems and their underlying dynamics is known, since the transition probabilities are governed by a coherent superposition of an exponentially growing number of transition amplitudes \cite{Aaronson-CC-2011,Tichy-II-2014}. Thus, the suppression laws in HCs suit the certification of many-particle indistinguishability in large systems as, e.g., required for the verification of any candidate boson-sampling device \cite{Spagnolo-EV-2014,Carolan-OE-2014,Tichy-SE-2014,Walschaers-SB-2016}. In realistic experimental scenarios particles will be only partially indistinguishable and the unitary will deviate from the ideal structure. Clearly, the suppression laws~(\ref{eq:supplaw1}) and (\ref{eq:supplaw2}) do not strictly hold in this regime. However, analogously to the procedure outlined for the Fourier suppression law in \cite{Tichy-SE-2014} the influence of both imperfections can be estimated: The deviation from the ideal case due to particle distinguishability is bounded by a function varying monotonously between zero for indistinguishable particles and the suppression ratio (see equations~(\ref{eq:suppratioBosons}) and (\ref{eq:suppratioFermions})) for fully distinguishable particles. For imperfect unitaries one can show that the probability of a suppressed event scales quadratically with the average relative deviation of the unitary. Ultimately, a precise calculation of the resulting probabilities is possible using the formalisms for partially distinguishable particles introduced in \cite{Shchesnovich-PI-2015,Tichy-SP-2015,Tillmann-GM-2015}. The question whether these formalisms can also benefit from symmetries remains open for future investigations.

Going beyond the single-particle regime may further be useful for applications in quantum information: Just as two-particle interference lies at the heart of two-qubit gates in photonic quantum computing \cite{OBrien-OQ-2007}, many-particle interference might enhance existing or enable new types of quantum algorithms on the HC. 

For realisations of high-dimensional HCs one can make use of long-range connections \cite{Daqing-DS-2011,Jukic-FD-2013} or internal degrees of freedom \cite{Boada-QS-2012}, even if only fewer dimensions are available in the embedding physical space. Particularly the latter approach has proven fruitful in recent years, with experiments successfully embedding $d=2$ dynamics in a $d=1$ atomic lattice \cite{Mancini-OC-2015,Stuhl-VE-2015} and specific proposals being made for reducing the required physical dimension by 1 for optomechanical ($d=3$) \cite{Schmidt-OC-2015} and optical systems ($d=4$) \cite{Ozawa-SD-2015}. Another promising route is offered by multi-mode-interference in optical fibers, potentially supporting graphs with highly-dimensional connectivity \cite{Defienne-TP-2016}. Besides a direct realisation of high-dimensional graph structures, the HC unitary for a fixed evolution time can also be implemented via cascaded directional couplers \cite{Reck-ER-1994} for which sophisticated optical platforms exist \cite{Carolan-UL-2015,Mower-HF-2015}, which may even be suitable for an observation of both bosonic and fermionic statistics on the same platform \cite{Matthews-OF-2013}.

%==================================================================================
%================================ACKNOWLEDGMENTS===================================
%==================================================================================
\ack

The authors would like to thank Andreas Buchleitner and Stefan Fischer for fruitful discussions and acknowledge support by the Austrian Science Fund (FWF projects I 2562-N27 and M 1849) and the Canadian Institute for Advanced Research (CIFAR, Quantum Information Science Program).

%==================================================================================
%================================APPENDIX==========================================
%==================================================================================

%==================================================================================
%==========================UNITARY ELEMENTS========================================
%==================================================================================
\appendix
\section{Derivation of the unitary elements}\label{aqq:Uelements}
\setcounter{section}{1}
The derivation of the suppression laws is based on the expression of unitary elements according to~(\ref{eq:Uelement}). This is obtained by considering the accumulation of $\pi/2$ phase terms for elements $\hat{U}_{j,1}$ due to subsequent applications of tensor powers in~(\ref{eq:U}), which leads to
\begin{eqnarray}\label{eq:Uderiv1}
\hat{U}_{j,1}=\frac{1}{\sqrt{n}}\exp\left(\rmi \frac{\pi}{4}\sum_{l=1}^d\left[1-x(j,2^l)\right]\right).
\end{eqnarray}
Analogously, additional phase terms are taken into account for columns $k$,
\begin{eqnarray}\label{eq:Uderiv2}
\hat{U}_{j,k}=\hat{U}_{j,1}\exp\left(\rmi \frac{\pi}{4}\sum_{m=1}^d\left[1-x(k,2^m)\right]x(j,2^m)\right).
\end{eqnarray}
Plugging~(\ref{eq:Uderiv1}) into~(\ref{eq:Uderiv2}) results in~(\ref{eq:Uelement}) after some algebraic steps.

%==================================================================================
%==================================================================================
%========================PROOF OF HC SUPP LAW======================================
%==================================================================================
%==================================================================================

\section{Proofs of the hypercube suppression laws}\label{aqq:SupLawProof}
In a fixed frame, the symmetry operations~(\ref{eq:S}) simply relabel the mode numbers, which motivates the introduction of operators $\mathcal{S}_\mathrm{d}(\bi{p})$, acting on mode numbers according to
\begin{eqnarray}\label{eq:Smas}
\mathcal{S}_\mathrm{d}(\bi{p})\ d_j(\bi{r})=d_j(\bi{r})+\sum_{k=1}^{\norm{\bi{p}}}\ x(d_j(\bi{r}),p_k)\frac{n}{p_k}.
\end{eqnarray}
Note, that the action on elements of mode assignment lists is indicated by the subscript~d. An important characteristic of the Rademacher functions~(\ref{eq:x}),
\begin{eqnarray}\label{eq:xsym}
x(d_j(\bi{r}),p')=\cases{
-x\left(\mathcal{S}_\mathrm{d}(\bi{p})\ d_j(\bi{r}),p'\right)&for $p'\in\bi{p}$\\\hspace{0.3cm} x\left(\mathcal{S}_\mathrm{d}(\bi{p})\ d_j(\bi{r}),p'\right)&for $p'\notin\bi{p}$,\\}
\end{eqnarray}
enables the expression of symmetric connections between unitary elements in~(\ref{eq:Uelement}),
\begin{eqnarray}\label{eq:Usym}
\hat{U}_{\mathcal{S}_\mathrm{d}(\bi{p})d_j(\bi{r}),d_k(\bi{s})}=\hat{U}_{d_j(\bi{r}),d_k(\bi{s})}\ \exp\left(\rmi\frac{\pi}{2}\phi(d_j(\bi{r}),d_k(\bi{s}),\bi{p})\right),
\end{eqnarray}
where the phase factor
\begin{eqnarray}\label{eq:Phi}
\phi(d_j(\bi{r}),d_k(\bi{s}),\bi{p})=\sum_{m=1}^{\norm{\bi{p}}}\ x(d_j(\bi{r}),p_m)\ x(d_k(\bi{s}),p_m)
\end{eqnarray}
is defined for brevity. Furthermore, using~(\ref{eq:Smas}), the symmetry condition~(\ref{eq:invariant}) for initial states can be expressed in terms of entries of the mode assignment list. Since the transition probability~(\ref{eq:P}) is independent on the ordering of $\bi{d}(\bi{r})$, it can be chosen conveniently, so that 
\begin{eqnarray}\label{eq:dsym}
d_{N/2+j}(\bi{r})=\mathcal{S}_\mathrm{d}(\bi{p})\ d_j(\bi{r})
\end{eqnarray}
for $j=1,\dots,N/2$.

%==================================================================================
%===============================BOSONS=============================================
%==================================================================================

\subsection{Bosons}
Considering $\sgn_{\mathrm{B}}(\bsigma)=1$ for bosonic particles and plugging the symmetry conditions~(\ref{eq:dsym}) and~(\ref{eq:Usym}) into~(\ref{eq:P}), the transition probability becomes
\begin{eqnarray}\label{eq:Psymderiv}
\fl
P_\mathrm{B}(\bi{r},\bi{s},\hat{U})\propto\left|\sum_{\bsigma\in S_{\bi{d}(\bi{s})}}\prod_{j=1}^{N/2} \hat{U}_{d_j(\bi{r}),\sigma_j} \hat{U}_{d_j(\bi{r}),\sigma_{N/2+j}} \exp\left(\rmi\frac{\pi}{2}\phi(d_j(\vec{r}),\sigma_{N/2+j},\bi{p})\right) \right|^2.
\end{eqnarray}
As the summation runs over all permutations $\bsigma$ of the final mode assignment list, these permutations are either symmetric with respect to order reversal, $\sigma_j=\sigma_{N/2+j}$, or possess a reversed-order partner $\bsigma'$, satisfying $\sigma'_j=\sigma_{N/2+j}$ and $\sigma_j=\sigma'_{N/2+j}$ for $j=1,\dots,N/2$. In the following, the latter case is considered first, while in the end it turns out, that final states which are covered by the suppression law cannot form symmetric permutations, such that the former case does not apply.  
The contribution of each pair $\bsigma$ and $\bsigma'$ in~(\ref{eq:Psymderiv}) then yields
\begin{eqnarray}
\propto& \prod_{j=1}^{N/2}\exp\left(\rmi\frac{\pi}{2}\phi(d_j(\bi{r}),\sigma_{N/2+j},\bi{p})\right)+\prod_{k=1}^{N/2}\exp\left(\rmi\frac{\pi}{2}\phi(d_k(\bi{r}),\sigma_k,\bi{p})\right)\\
\label{eq:contributionBoson}
\propto& 1+\prod_{j=1}^{N/2}\exp\left(\rmi\frac{\pi}{2}\left[\phi(d_j(\bi{r}),\sigma_{N/2+j},\bi{p})-\phi(d_j(\bi{r}),\sigma_j,\bi{p})\right]\right)
\end{eqnarray}
which vanishes if 
\begin{eqnarray}\label{eq:mod01}
\mod\left[\sum_{j=1}^{N/2}\ \phi(d_j(\bi{r}),\sigma_{N/2+j},\bi{p})-\phi(d_j(\bi{r}),\sigma_j,\bi{p})\ ,\ 4\right]=2.
\end{eqnarray}
The condition for final states fulfilling~(\ref{eq:mod01}) can be shown by defining the sets
\begin{eqnarray}\label{eq:Pset}
P(d_j(\bi{r}))&=\{p_m\ |\ p_m\in\bi{p}\ \wedge\ x(d_j(\bi{r}),p_m)=1\},\\
\bar{P}(d_j(\bi{r}))&=\{p_{\bar{m}}\ |\ p_{\bar{m}}\in\bi{p}\ \wedge\ x(d_j(\bi{r}),p_{\bar{m}})=-1\}
\end{eqnarray}
with which the phase factors in~(\ref{eq:mod01}) can be re-written to
\begin{eqnarray}
\phi(d_j(\bi{r}),\sigma_k,\bi{p})
&=\sum_{p\in P(d_j(\bi{r}))}\ x(\sigma_k,p)-\sum_{\bar{p}\in\bar{P}(d_j(\bi{r}))}\ x(\sigma_k,\bar{p})\\
&=\sum_{m=1}^{\norm{\bi{p}}}\ x(\sigma_k,p_m)-2\sum_{\bar{p}\in\bar{P}(d_j(\bi{r}))}\ x(\sigma_k,\bar{p}).\label{eq:secondsum}
\end{eqnarray}
for any $k\in\{1,\dots,N\}$. The second term in~(\ref{eq:secondsum}) yields
\begin{eqnarray}
\mod\left[2\sum_{\bar{p}\in\bar{P}(d_j(\bi{r}))}\ x(\sigma,\bar{p})\ ,\ 4\right]=\cases{
0&for even $\norm{\bar{P}(d_j(\bi{r}))}$ \\
2&for odd $\norm{\bar{P}(d_j(\bi{r}))}$,\\}
\end{eqnarray}
and does not contribute to the remainder in~(\ref{eq:mod01}) since
\begin{eqnarray}\label{eq:resultPset}
\mod\left[2\sum_{\bar{p}\in\bar{P}(d_j(\bi{r}))}\ x(\sigma_j,\bar{p})-x(\sigma_{N/2+j},\bar{p})\ ,\ 4\right]=0.
\end{eqnarray}
Condition~(\ref{eq:mod01}) then becomes
\begin{eqnarray}\label{eq:mod02}
\mod\left[\sum_{j=1}^{N/2}\sum_{m=1}^{\norm{\bi{p}}}\ x(\sigma_{N/2+j},p_m)-x(\sigma_j,p_m)\ ,\ 4\right]=2.
\end{eqnarray}
In order to rewrite the sums over all elements in $\bi{p}$, we utilize the identities
\begin{eqnarray}\label{eq:identity1}
\mod\left[\sum_{m=1}^{\norm{\bi{p}}}\ x(j,p_m)\ ,\ 4\right]=(-1)^{(\norm{\bi{p}}+2)/2} \prod_{m=1}^{\norm{\bi{p}}}\ x(j,p_m)+1
\end{eqnarray}
for even $\norm{\bi{p}}$ and
\begin{eqnarray}\label{eq:identity2}
\mod\left[1+\sum_{m=1}^{\norm{\bi{p}}}\ x(j,p_m)\ ,\ 4\right]=(-1)^{(\norm{\bi{p}}-1)/2} \prod_{m=1}^{\norm{\bi{p}}}\ x(j,p_m)+1
\end{eqnarray}
for odd $\norm{\bi{p}}$. These identities can be proven by induction on $\norm{\bi{p}}$. Then, for both cases,~(\ref{eq:identity1}) and~(\ref{eq:identity2}), condition~(\ref{eq:mod02}) can be rewritten, using the Walsh functions~(\ref{eq:partition}):
\begin{eqnarray}\label{eq:mod03}
\mod\left[\sum_{j=1}^{N/2}\mathcal{A}(\sigma_{N/2+j},\bi{p})-\mathcal{A}(\sigma_{j},\bi{p})\ ,\ 4\right]=2.
\end{eqnarray}
Further, we use the identity
\begin{eqnarray}
\mod\left[\sum_{j=1}^{N/2}\mathcal{A}(\sigma_{N/2+j},\bi{p})-\mathcal{A}(\sigma_{j},\bi{p})\ ,\ 4\right]=1-\prod_{j=1}^{N}\mathcal{A}(\sigma_j,\bi{p}),
\end{eqnarray}
which can be proven by induction on $N$, such that condition~(\ref{eq:mod03}) becomes
\begin{eqnarray}
\prod_{j=1}^{N}\mathcal{A}(\sigma_j,\bi{p})=\prod_{j=1}^{N}\mathcal{A}(d_j(\bi{s}),\bi{p})=-1.
\end{eqnarray}
Since $N$ is even, this is the case if the final state $\bi{d}(\bi{s})$ contains an odd number of particles in the subset of modes $k$ for which $\mathcal{A}(k,\bi{p})=1$. Clearly, those final states $\bi{d}(\bi{s})$ can't form symmetric permutations adhering to $\sigma_j=\sigma_{N/2+j}$, which had to be shown.

%==================================================================================
%=================================FERMIONS=========================================
%==================================================================================

\subsection{Fermions}

For fermions, the sign factor $\sgn_\mathrm{F}(\bsigma)=\sgn(\bsigma)$ needs to be considered in the calculation of transition probabilities according to~(\ref{eq:P}). By defining
\begin{eqnarray}
J(q)=\left\{j\ |\ j\in\{1,\dots,N\}\setminus \{N-q,N/2-q\}\right\}
\end{eqnarray}
for $q\in\{0,\dots,N/2-1\}$, each permutation $\bsigma$ can be assigned a partner permutation $\bsigma^{(q)}$, for which the elements $N/2-q$ and $N-q$ are exchanged,
\begin{eqnarray}\label{eq:sigmatilde}
\sigma^{(q)}_j=\cases{
\sigma_j&for $j\in J(q)$\\
\sigma_{N-q}&for $j=N/2-q$\\
\sigma_{N/2-q}&for $j=N-q$,\\}
\end{eqnarray}
and accordingly $\sgn(\bsigma)=-\sgn(\bsigma^{(q)})$. Then, the summands of $\bsigma$ and $\bsigma^{(q)}$ in~(\ref{eq:P}) yield
\begin{eqnarray}
&\sgn(\bsigma)\left(\prod_{j=1}^N \hat{U}_{d_j(\bi{r}),\sigma_j}-\prod_{k=1}^N \hat{U}_{d_k(\bi{r}),\sigma^{(q)}_k}\right)\\
=&\sgn(\bsigma)\left(\prod_{j=1}^N \hat{U}_{d_j(\bi{r}),\sigma_j}-\prod_{k\in J(q)}\hat{U}_{d_k(\bi{r}),\sigma_k}\ \hat{U}_{d_{\frac{N}{2}-q}(\bi{r}),\sigma_{N-q}} \hat{U}_{d_{N-q}(\bi{r}),\sigma_{\frac{N}{2}-q}}\right)\\
\propto& 1-\exp\left(\rmi\frac{\pi}{2}\left[\phi(d_{N-q}(\bi{r}),\sigma_{N-q},\bi{p})+\phi(d_{N/2-q}(\bi{r}),\sigma_{N/2-q},\bi{p})\right]\right)
\end{eqnarray}
where we used~(\ref{eq:dsym}) and (\ref{eq:Usym}) in the last step. This contribution vanishes if
\begin{eqnarray}\label{eq:Fmod01}
\mod\left[\phi(d_{N-q}(\bi{r}),\sigma_{N-q},\bi{p})+\phi(d_{N/2-q}(\bi{r}),\sigma_{N/2-q},\bi{p})\ ,\ 4\right]=0.
\end{eqnarray}
Utilizing~(\ref{eq:xsym}) in definition~(\ref{eq:Phi}), condition~(\ref{eq:Fmod01}) becomes
\begin{eqnarray}
\mod\left[\phi(d_{N-q}(\bi{r}),\sigma_{N-q},\bi{p})-\phi(d_{N-q}(\bi{r}),\sigma_{N/2-q},\bi{p})\ ,\ 4\right]=0
\end{eqnarray}
which results in
\begin{eqnarray}\label{eq:Fcancelcond}
\mathcal{A}(\sigma_{N-q},\bi{p})=\mathcal{A}(\sigma_{N/2-q},\bi{p})
\end{eqnarray}
by proceeding as in~(\ref{eq:Pset})~-~(\ref{eq:mod03}). Thus, the summands for $\bsigma$ and $\bsigma^{(q)}$ cancel each other out if condition~(\ref{eq:Fcancelcond}) holds. 

For a set $\bi{q}=(q_1,q_2,\dots)$, where each $q$ value is contained at most once, we define the permutation $\bsigma^{(\bi{q})}$ as
\begin{eqnarray}
\sigma^{(\bi{q})}_j=\cases{
\sigma_j&for $j\in \bigcap\limits_{q\in\bi{q}}J(q)$\\
\sigma_{N-q}&for $j=N/2-q\ \forall\ q\in\bi{q}$ \\
\sigma_{N/2-q}&for $j=N-q\ \forall\ q\in\bi{q}$.\\}
\end{eqnarray}
Now, it is possible to group all $N!$ possible permutations of the final mode assignment list into sets $\{\bsigma^{(\bi{q}_1)},\bsigma^{(\bi{q}_2)},\dots \}$, containing $2^{N/2}$ permutations $\bsigma^{(\bi{q}_i)}$, where the sets $\bi{q}_i$ are given by all possible subsets of $\{0,1,\dots,N/2-1\}$ including the empty list. Accordingly, for every single $q$, all permutations of a set $\{\bsigma^{(\bi{q}_1)},\bsigma^{(\bi{q}_2)},\dots \}$ can be grouped into pairs, whose elements $N/2-q$ and $N-q$ are exchanged. Then, if condition~(\ref{eq:Fcancelcond}) is fulfilled for any single $q$ value, all permutations of the set $\{\bsigma^{(\bi{q}_1)},\bsigma^{(\bi{q}_2)},\dots \}$ cancel pairwise. Consequently, there are sets  $\{\bsigma^{(\bi{q}_1)},\bsigma^{(\bi{q}_2)},\dots \}$, in which no pairwise cancellation occurs if and only if $\mathcal{A}(\sigma_{j},\bi{p})=1$ for half of all $j\in\{1,\dots,N\}$ and $\mathcal{A}(\sigma_{j},\bi{p})=-1$ for the other half. Thus, all final states adhering
\begin{eqnarray}
\sum_{j=1}^N \mathcal{A}(d_j(\bi{s}),\bi{p})\neq 0
\end{eqnarray}
must be suppressed.

\section{Derivation of the generalization to arbitrary subunitaries}\label{app:GeneralProof}

In order to describe the effect of symmetry operations~(\ref{eq:SymGen}) on mode numbers, the same operators as in~(\ref{eq:Smas}) can be used, however, with the restriction to segmentations $p\in\{2,4,8,\dots,2^d\}$, for which the identities~(\ref{eq:xsym}) hold as well. The definition
\begin{eqnarray}
f(l,m)=1+\mod\left[l-1\ ,\ m\right]
\end{eqnarray}
for $l=1,\dots, 2^dm$ ascribes the index within the subunitary and satisfies
\begin{eqnarray}\label{eq:f}
f(\mathcal{S}_\mathrm{d}(\bi{p})d_j(\bi{r}),m)=f(d_j(\bi{r}),m).
\end{eqnarray}
Thus, the unitary elements in~(\ref{eq:unitarygen}) read
\begin{eqnarray}
\hat{\mathcal{U}}_{j,k}=\hat{A}_{f(j,m),f(k,m)}\frac{1}{\sqrt{2^d}}\exp\left(\rmi\frac{\pi}{4}\left[d-\sum_{l=1}^d\ x(j,2^l)\ x(k,2^l)\right]\right),
\end{eqnarray}
where $\hat{A}$ denotes the $m\times m$ subunitary. Expectedly, the dynamics of the subunitary $\hat{A}$ and the HC graph are decoupled. By exploiting~(\ref{eq:xsym}) and (\ref{eq:f}), the symmetric connection between unitary elements obeys
\begin{eqnarray}
\hat{\mathcal{U}}_{\mathcal{S}_\mathrm{d}(\bi{p})d_j(\bi{r}),d_k(\bi{s})}=\hat{\mathcal{U}}_{d_j(\bi{r}),d_k(\bi{s})}\ \exp\left(\rmi\frac{\pi}{2}\phi(d_j(\bi{r}),d_k(\bi{s}),\bi{p})\right).
\end{eqnarray}
Clearly, this is identical to the symmetry relation~(\ref{eq:Usym}). In conclusion, the same symmetry suppression laws hold for unitaries of the form~(\ref{eq:unitarygen}) with the restriction to segmentation values $p\in\{2,4,8,\dots,2^d\}$.

%==================================================================================
%================================REFERENCES========================================
%==================================================================================

\section*{References}
%Automatic bibliography, turn off for arxiv-Version
%\bibliography{CDbibHyperCube}

\end{document}